\documentclass[aps,prx,final,10pt,twocolumn,superscriptaddress]{revtex4-2}
\usepackage{amsmath,amssymb,bm}
\usepackage{graphicx,color}
\usepackage[squaren]{SIunits}
\usepackage[english]{babel}
\usepackage{amstext}
\usepackage{amsthm}
\usepackage{latexsym}
\usepackage{array}
\usepackage{color}
\usepackage{float}
\usepackage{microtype}
 
\usepackage{multirow}
\usepackage{dcolumn}
\usepackage{soul}
\usepackage{lettrine}
\usepackage{type1cm} 
\usepackage{natbib}
\usepackage{longtable}
\definecolor{url}{RGB}{0,20,160}
\usepackage[colorlinks=true,linkcolor=blue,citecolor=blue,urlcolor=url]{hyperref}
\usepackage[usenames,dvipsnames,svgnaes,table]{xcolor}
\makeatletter
\def\frutiger{cmss10 }
\def\frutigerbold{cmssbx10 }
=\frutigerbold at 15pt
=\frutiger at 10pt
=\frutigerbold at 12pt
=\frutigerbold at10pt
=\frutigerbold at 8pt
=\frutiger at 8pt
=\frutigerbold at 31pt
\def\@caption@tabnum@sep{\figtextfont{{ }{\bf\textbar}{ }}}%
\def\fnum@table{{\bf\tablename~\thetable}}

\def\@caption@fignum@sep{\figtextfont{{ }{\bf\textbar}{ }}}%
\def\fnum@figure{{\bf\figurename~\thefigure}}
\renewenvironment{figure}{\@float{figure}\def\textbf##1{{\fignumfont ##1}}\def\bf{\fignumfont}}{\end@float}
\def\@startsection#1#2#3#4#5#6{%
	\if@noskipsec\leavevmode\fi
	\par\@tempskipa #4\relax
	\@afterindenttrue
	\ifdim\@tempskipa <\z@
	\@tempskipa -\@tempskipa \@afterindentfalse
	\fi\if@nobreak\everypar{}%
	\else\addpenalty\@secpenalty\addvspace\@tempskipa\fi
	\@ifstar{\@ssect{#3}{#4}{#5}{#6}}{\@dblarg{\@sect{#1}{#2}{#3}{#4}{#5}{#6}}}}
\def\@sect#1#2#3#4#5#6[#7]#8{%
	\ifnum #2>0
	\let\@svsec\@empty
	\else\refstepcounter{#1}\protected@edef\@svsec{\@seccntformat{#1}\relax}\fi
	\@tempskipa #5\relax
	\ifdim\@tempskipa>\z@
	\begingroup#6{\@hangfrom{\hskip #3\relax\@svsec}%
		\interlinepenalty \@M #8\@@par}\endgroup
	\csname #1mark\endcsname{#7}%
	\addcontentsline{toc}{#1}{%
		\ifnum #2>\c@secnumdepth\else
		\protect\numberline{\csname the#1\endcsname}\fi #7}%
	\else\def\@svsechd{#6{\hskip #3\relax
			\@svsec #8\ifnum#2=2.\fi}%
		\csname #1mark\endcsname{#7}%
		\addcontentsline{toc}{#1}{%
			\ifnum #2>\c@secnumdepth \else
			\protect\numberline{\csname the#1\endcsname}\fi #7}}%
	\fi\@xsect{#5}}

\renewcommand\section{\@startsection {section}{1}{\z@}%
	{-10pt \@plus -1ex \@minus -.2ex}{.5ex }{\normalfont\Large\bfseries\sectionfont}}
\renewcommand\subsection{\@startsection{subsection}{2}{\z@}%
	{10pt\@plus 1ex \@minus .2ex}{-0.5ex \@plus .2ex}{\normalfont\large\bfseries\subsectionfont}}
\def\frontmatter@title@format{\titlefont\centering}%
\def\frontmatter@title@below{\addvspace{-5pt}}%

\renewcommand\NAT@biblabelnum[1]{#1.}
\renewcommand\NAT@citesuper[3]{\ifNAT@swa
	\unskip\hspace{1\p@}\textsuperscript{(#1)}%
	\if\relax#3\relax\else\ (#3)\fi\else (#1)\fi\endgroup}

\newcommand*\bib@heading{%
	\section{\refname}
	\fontsize{8}{10}\selectfont
}
\newcommand*\@openbib@code{%
	\advance\leftmargin\bibindent
	\itemindent -\bibindent
	\listparindent \itemindent
	\parsep \z@
}%
\newdimen\bibindent
\usepackage[usenames,dvipsnames,svgnaes,table]{xcolor}
\definecolor{col1}{rgb}{0.0, 0.30, 1.0}
\definecolor{col2}{rgb}{0.9, 0.0, 0.30}

\usepackage[draft]{changes} 
\usepackage[normalem]{ulem}
\definechangesauthor[name={yi}, color=red]{Yi}

\makeatletter

\newcommand{\Rmnum}[1]{\expandafter\@slowromancap\romannumeral #1@}
\newcommand{\revise}[1]{{\textcolor{red}}}

\begin{document}
\title{Ultrastrong coupling between polar distortion and optical properties in ferroelectric MoBr$_2$O$_2$}

\author{Zhaojun Li}
\affiliation{Key Laboratory of Advanced Materials and Devices for Post-Moore Chips, Ministry of Education, University of Science and Technology Beijing, Beijing 100083, China}
\affiliation{School of Mathematics and Physics, University of Science and Technology Beijing, Beijing 100083, China}

\author{Lorenzo Varrassi}
\affiliation{Department of Physics and Astronomy ”Augusto Righi”, Alma Mater Studiorum - Universit\`{a} di Bologna, 40127 Bologna, Italy}

\author{Yali Yang}
\email{ylyang@ustb.edu.cn}
\affiliation{School of Mathematics and Physics, University of Science and Technology Beijing, Beijing 100083, China}

\author{Cesare Franchini}
\affiliation{Faculty of Physics and Center for Computational Materials Science, University of Vienna, Kolingasse 14-16, 1090 Vienna, Austria}
\affiliation{Department of Physics and Astronomy ”Augusto Righi”, Alma Mater Studiorum - Universit\`{a} di Bologna, 40127 Bologna, Italy}

\author{Laurent Bellaiche}
\affiliation{Physics Department and Institute for Nanoscience and Engineering, University of Arkansas, Fayetteville, Arkansas 72701, USA}

\author{Jiangang He}
\email{jghe2021@ustb.edu.cn}
\affiliation{Key Laboratory of Advanced Materials and Devices for Post-Moore Chips, Ministry of Education, University of Science and Technology Beijing, Beijing 100083, China}
\affiliation{School of Mathematics and Physics, University of Science and Technology Beijing, Beijing 100083, China}

\date{\today}

\begin{abstract}
\noindent 
Tuning the properties of materials using external stimuli is crucial for developing versatile smart materials. A strong coupling among order parameters within a single-phase material constitutes a potent foundation for achieving precise property control. However, cross-coupling is pretty weak in most single materials. Leveraging first principles calculations, we demonstrate the layered mixed anion compound MoBr$_2$O$_2$ exhibits electric-field switchable spontaneous polarization and ultrastrong coupling between polar distortion and electronic structures as well as optical properties. It offers feasible avenues of achieving tunable Rashba spin-splitting, electrochromism, thermochromism, photochromism, and nonlinear optics by applying an external electric field to a single domain sample, heating, as well as intense light illumination. Additionally, it exhibits an exceptionally large photostrictive effect. These findings not only showcase the feasibility of achieving multiple order parameter coupling within a single material, but also pave the way for comprehensive applications based on property control, such as energy harvesting, information processing, and ultrafast control.
\end{abstract}

\maketitle

Coupling among multiple order parameters in a single-phase material offers great opportunities to tune the physical properties of the material using external stimuli, making it of significant interest in both fundamental research and technological applications~\cite{asamitsu1997current,PhysRevB.61.594,he2018tunable,PhysRevLett.123.096801}. Since the crystal structure is of paramount importance in determining the properties of materials, any perturbation to the structure directly influences the modification of the key characteristics, such spin~\cite{PhysRevLett.97.267602,forst2011nonlinear}, orbitals~\cite{PhysRevLett.101.197404}, and charge~\cite{PhysRevLett.112.157002}. Ferroelectric materials are comprised of polar structure with the direction and magnitude of spontaneous electric polarization switchable by an external electric field, exhibiting strong coupling between polarization and other order parameters, such as lattice strain in ferroelectrics~\cite{PhysRevLett.17.198,kundys2010light} and magnetic field in magnetoelectric multiferroics~\cite{fiebig2005revival,10.1063/1.2836410}. More importantly, the switchability of ferroelectrics enables to tune the properties of materials that are strongly coupled with changes of the polar structure. For example, the switchable Rashba spin splitting in GeTe was initially predicted and subsequently experimentally verified~\cite{https://doi.org/10.1002/adma.201203199,doi:10.1021/acs.nanolett.7b04829,liebmann2016giant}. An organic molecular ferroelectric is found to show light-triggered structural change with reversible photoisomerization~\cite{https://doi.org/10.1002/advs.202102614}. Reversible lattice strain, also known as photostrictive effect~\cite{10.1063/1.4905505}, has been observed in several ferroelectric materials because of the lattice strain associated with polarization change~\cite{PhysRevLett.17.198,kundys2010light,PhysRevB.96.045205,PhysRevLett.123.087601}.

In addition to the conventional method of switching polarization using an external electric field, which entails circuitry access and has a sluggish switching time, electromagnetic waves can also be utilized to switch spontaneous polarization as well~\cite{https://doi.org/10.1002/adom.202002146}. So far, several approaches have been developed to manipulate polarization in ferroelectric materials using laser pulses:\cite{PhysRevB.106.L140302,chen2022deterministic} nonlinear coupling between the soft mode phonons and the near-infrared pluses via impulsive stimulated Raman scattering~\cite{10.1063/1.453733,PhysRevLett.73.1122}, resonant coupling between the soft mode phonons and the terahertz pulse~\cite{PhysRevLett.108.097401,PhysRevB.94.180104,doi:10.1126/science.aaw4913}, indirect manipulation of soft mode phonons by midinfrared pulses~\cite{PhysRevLett.118.197601,Imbrock:20}, and photovoltaic effect~\cite{PhysRevLett.108.087601,PhysRevLett.112.097602,PhysRevLett.108.087601}. In particular, intense THz pulses can rapidly switch the polarization of ferroelectric materials~\cite{PhysRevLett.118.197601,PhysRevB.94.180104,https://doi.org/10.1002/adma.200802094}. The advantage of ultrashort femtosecond laser pulses is their ability to generate high peak electric fields (over 100 MVcm$^{-1}$)~\cite{koulouklidis2020observation}. The carriers generated by above-bandgap light illumination have the ability to screen the spontaneous polarization of ferroelectric materials. As a result, the ferroelectric phase is destabilized by suppressing its polar distortion, similar to the way that the doped carriers screen the polarization of ferroelectric materials~\cite{PhysRevLett.104.147602,PhysRevLett.109.247601,PhysRevB.96.045205,PhysRevLett.118.227401,PhysRevLett.123.087601}. Although this method is unable to switch the direction of polarization, it is possible to achieve the non-polar reference state under intense light illumination, which is similar to the case of prompting ferroelectric phase transition by elevating temperature. All these phenomena are based on manipulating the polar structural distortion of ferroelectric materials and rely on the strength of coupling among order parameters.

However, the coupling among order parameters is usually very weak, which significantly limits the effectiveness of cross-control and hinders the development of smart materials. In this work, ultrastrong couplings have been discovered between the polar distortion and the electronic structure, as well as between the polar distortion and the optical properties in MoBr$_2$O$_2$ through density functional theory calculations. The bandgap, optical absorption, birefringence, and lattice constants of MoBr$_2$O$_2$ exhibit a high degree of coupling with the polar distortion of the ground state phase. Moreover, we demonstrate the spontaneous polarization of MoBr$_2$O$_2$ is switchable by applying an external electric field. Therefore, tunable electrochromism, thermochromism, photochromism, and nonlinear optics can be achieved in MoBr$_2$O$_2$ bulk by applying an external electric field, heating, and intense light illumination. Additionally, an exceptionally large photostrictive effect is observed, which is 27 times larger than BiFeO$_3$, 6 times larger than the archetypal $d^0$ ferroelectric materials BaTiO$_3$ (4 times of PbTiO$_3$)~\cite{PhysRevB.96.045205}, and 2 times larger than that of two-dimensional monochalcogenide compounds~\cite{PhysRevLett.118.227401}. These results not only demonstrate the feasibility of achieving robust coupling between multiple order parameters in bulk materials but also pave the way for manipulating the properties of materials using external stimuli, showing promising applications in energy harvesting, ultrafast control, and information processing.

\begin{figure}
	\centering
	\includegraphics[width=1.0\linewidth]{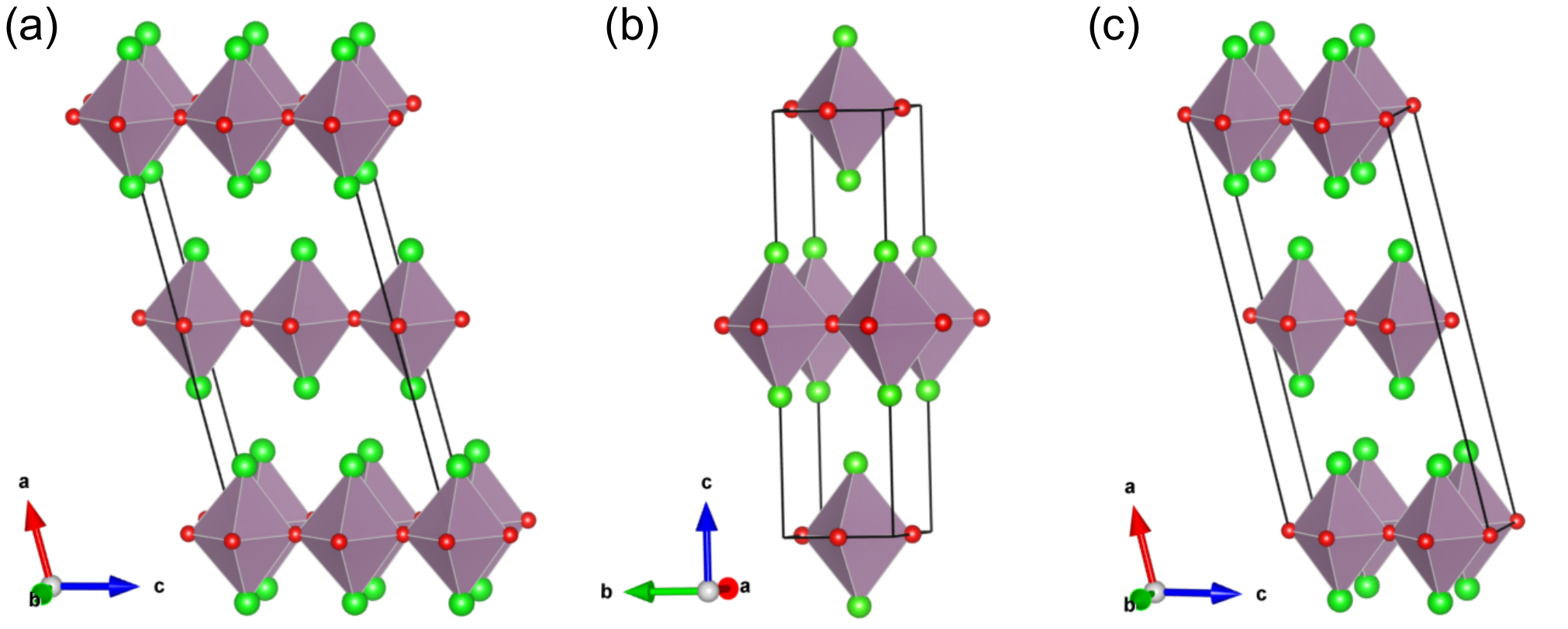}
	\caption{Crystal structures of MoBr$_2$O$_2$. (a) $Cc$ (b) $I4/mmm$ (c) $C2/c$. Large (green) and small (red) spheres refer to Br and O, respectively. The center of orchid octahedral are Mo atoms.}
	\label{crystalstructure}
\end{figure} 

\section{Results and discussion}
\noindent \textbf{Ground-state structure and phase transitions} \\
\noindent MoBr$_2$O$_2$ crystallizes in a polar space group ($Cc$, No. 9) in experiment~\cite{https://doi.org/10.1002/zaac.201100042}. The crystal structure consists of corner-shared MoBr$_2$O$_4$ octahedral aligned in the $b$ and $c$ directions, while van der Waals gap separates the MoBr$_2$O$_4$ octahedral along the $a$ direction, see Figure~\ref{crystalstructure}(a). The heteroleptic MoBr$_2$O$_4$ octahedron exhibits ordering of two anions, O$^{2-}$ and Br$^{-}$. Specifically, there are four O$^{2-}$ anions situated in the equatorial positions, while two Br$^{-}$ anions occupy the pole positions. Therefore, all O$^{2-}$ anions are shared by two MoBr$_2$O$_4$ octahedral, while every Br$^{-}$ only bounds with one octahedron, forming a van der Waals layered structure. Several other members of $MX_2$O$_2$ ($M$=Mo and W; $X$=Cl and Br) family have also been synthesized. Recently, MoCl$_2$O$_2$ was found to have a structure with the space group $Fmm2$ (No. 42)~\cite{https://doi.org/10.1002/anie.202310835}. An earlier study determined that WCl$_2$O$_2$ adopts the $P2_1am$ (No. 26) space group based on single-crystal X-ray diffraction~\cite{jarchow1968kristallstruktur}. However, a subsequent powder neutron diffraction study by Abrahams {\it et al.} identified it as a Cl$^{-}$ and O$^{2-}$ disordered occupied structure with the space group $Immm$ (No. 74)~\cite{abrahams1993disordered}. All experimental measurements were conducted at room temperature, and no phase transition has been reported for any of these compounds. A summary of the known $MX_2$O$_2$ compounds reported in experiment is tabulated in Table~\textcolor{magenta}{S1} of the supporting information. Our fully relaxed lattice constants of MoBr$_2$O$_2$ at 0 K are $a$=15.1901 \AA, $b$=3.8889 \AA, and $c$=7.6133 \AA, $\beta$=104.467$^\circ$. These values agree well with the experimental values (15.2233 \AA, 3.9061 \AA, 7.7109 \AA, $\beta$=104.394$^\circ$) measured at room temperature~\cite{https://doi.org/10.1002/zaac.201100042}, taking into account the thermal expansion that occurs at finite temperatures.

\begin{figure}
	\centering
	\includegraphics[width=1.0\linewidth]{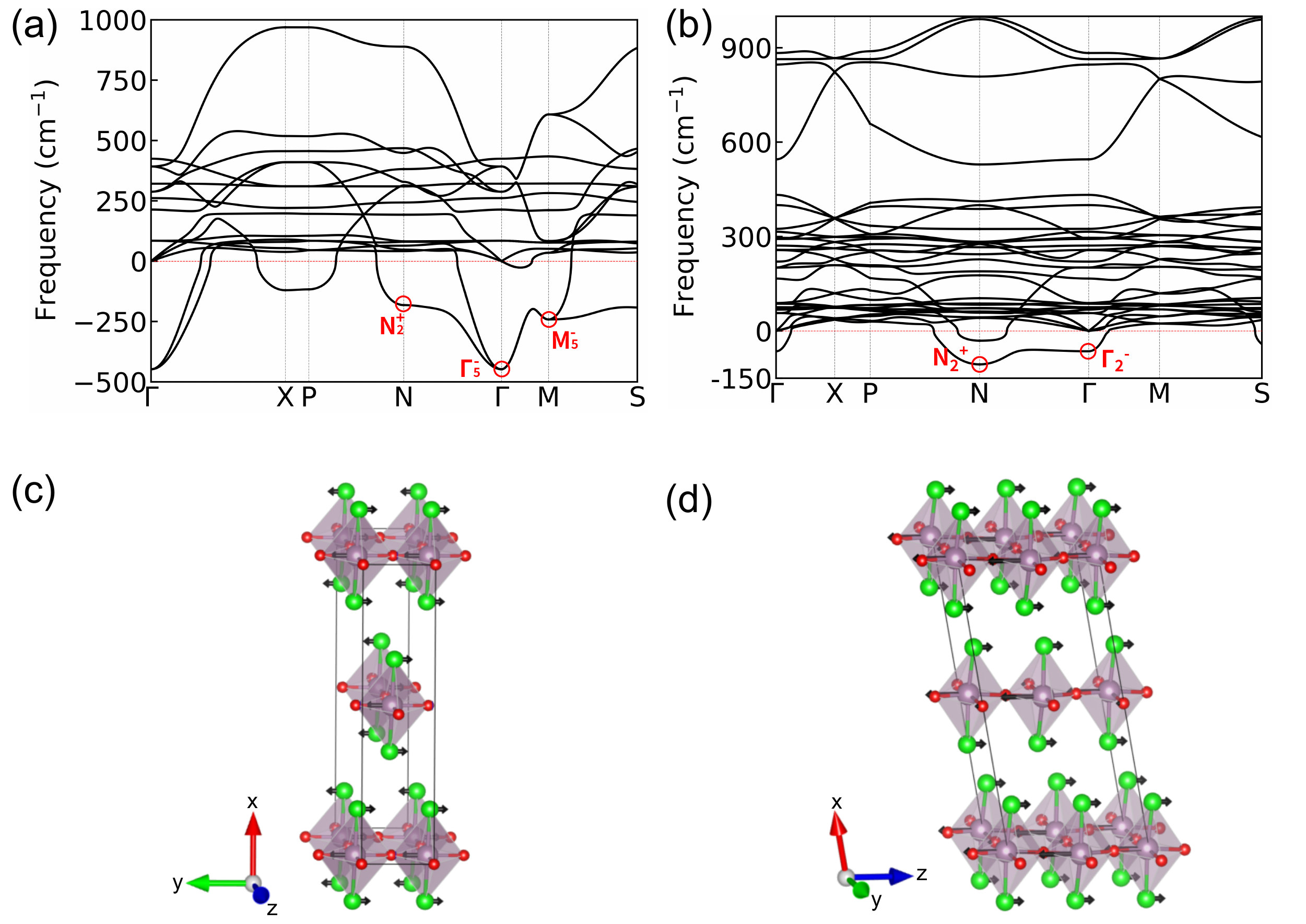}
	\caption{Phonon dispersion and real space atomic displacements associated with imaginary frequencies of MoBr$_2$O$_2$. (a) and (b) are the phonon dispersion of $I4/mmm$ and $C2/c$ MoBr$_2$O$_2$ calculated using SCAN+rvv10. (c) and (d) are the real space atomic displacements of $N_2^+$ of $I4/mmm$ and $\Gamma_2^-$ of $C2/c$, respectively.}
	\label{phononi4mmm}
\end{figure} 

With a nominal oxidation state of 6+ and a $d^{0}$ configuration, $M^{6+}$ cation strongly tends to lift orbital degeneracy in $MX_2$O$_4$ octahedral through structure deformation, which is commonly referred to as the pseudo Jahn-Teller effect~\cite{doi:10.1021/acs.chemrev.0c00718}. Therefore, it is likely that all the experimentally observed phases are the consequences of $MX_2$O$_4$ octahedral distortions. To comprehend the potential energy landscape and ascertain the root cause of the experimentally observed $Cc$ structure, we calculate the phonon spectra of the supergroup structures and employ isotropy group theory~\cite{isotropy} to analyze the unstable phonon modes. This approach enables us to gain insights into the specific atomic displacements that contribute to the structural distortion, unraveling the underlying dynamics and energetics of the system. The highest symmetry structure of $Cc$ and $Fmm2$ is $I4/mmm$ (see Figure~\ref{crystalstructure}), where four equatorial $M$-O bonds in $MX_2$O$_4$ octahedral are equal and Wyckoff position of $M$ site possesses D$_{\rm 4h}$ point group. As shown in Figure~\textcolor{magenta}{S1} of supporting information, $Cc$ phase is formed by combining $N_2^+$ along the (1,0,0,0) direction of the order parameter connected with the isotropy subgroup and $\Gamma_5^-$ along the (1,0) direction of the order parameter of the parent phase $I4/mmm$~\cite{Stokes:pd5086}, while $Fmm2$ can be generated solely by $\Gamma_5^-$ along (1,1) direction. The phonon dispersion of $I4/mmm$ MoBr$_2$O$_2$ is shown in Figure~\ref{phononi4mmm} and the other compounds are shown in Figure~\textcolor{magenta}{S2} of the supporting information. Indeed, the primary unstable phonon modes observed in all these compounds are $N_2^+$ and $\Gamma_5^-$, although their magnitudes decrease as $M$ transitions from Mo to W and $X$ from Cl to Br and finally to I. The phonon calculations unequivocally support the symmetry analysis based on the isotropy subgroup theory~\cite{Stokes:pd5086}.

\begin{figure*}
	\centering
	\includegraphics[width=1.0\linewidth]{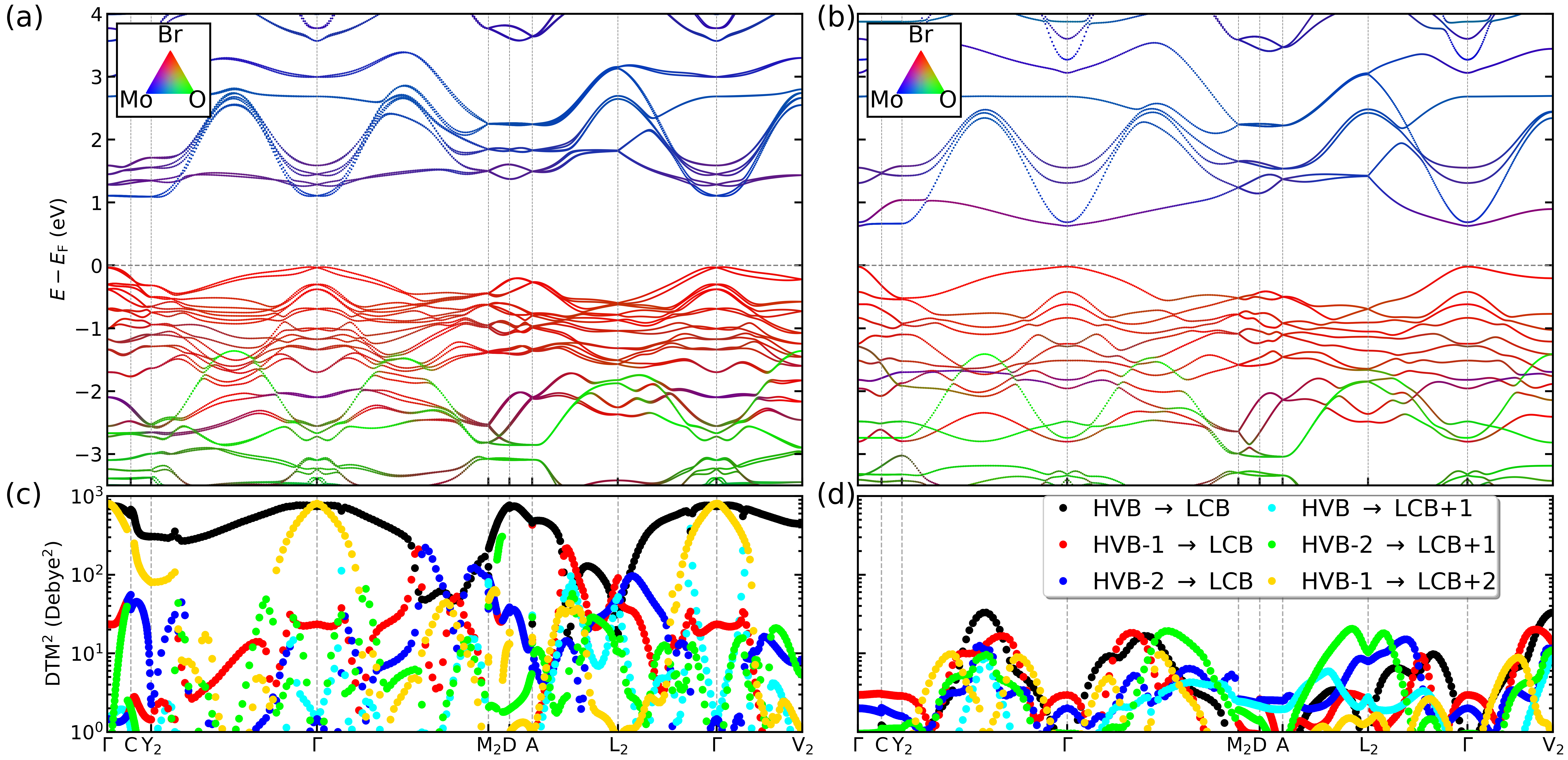}
	\caption{Band structures and dipole transition matrix of MoBr$_2$O$_2$ calculated using PBE functional. (a) and (b) are the band structures of $Cc$ and $C2/c$ phases, respectively, including spin-orbital-coupling (SOC). Blue, red, and green colors represent the contributions of Mo 4$d$, Br 4$p$, and O 2$p$ orbitals to a band, respectively. (c) and (d) are the squares of the dipole transition matrix elements (DTM$^2$) for $Cc$ and $C2/c$ phases, respectively, without including SOC.}
	\label{bandstructure}
\end{figure*} 

As shown in Figure~\ref{phononi4mmm}(c), the unstable phonon mode $N_2^{+}$ along the (1,0,0,0) direction of the order parameter connected with the isotropy subgroup corresponds to an antipolar distortion, where the Mo and Br ions in two adjacent MoBr$_2$O$_4$ octahedral move along opposite directions in the $z$ direction. However, in the $y$ direction, the displacements of Mo and Br ions in two adjacent MoBr$_2$O$_4$ octahedral are in the same direction. The displacements of Mo and Br ions within one MoBr$_2$O$_4$ octahedron is reminiscent of the polar distortion observed in the archetypal $d^0$ ferroelectric BaTiO$_3$~\cite{PhysRevB.60.836}, due to the pseudo Jahn-Teller effect~\cite{doi:10.1021/acs.chemrev.0c00718}. The condensation of $N_2^+$ mode along (1,0,0,0) direction leads to the $C2/c$ structure, see Figure~\ref{crystalstructure}(c). The $C2/c$ exhibits an unstable phonon mode at the $\Gamma$ point, {\it i.e.,} $\Gamma_2^-$ mode, which is connected to the $Cc$ through a unique direction, see Figure~\ref{phononi4mmm}(b). The $\Gamma_2^-$ mode of $C2/c$ phase is polar, with the main ion displacement (Mo$^{6+}$ and Br$^{-}$) along the $z$ direction, which is perpendicular to that of $N_2^+$ mode. The energy gains by successive condensing $N_2^+$ (from $I4/mmm$ to $C2/c$) and $\Gamma_2^-$ (from $C2/c$ to $Cc$) are -38 and -11 meV/atom, respectively. Therefore, it is likely that the polarization of the $Cc$ phase can be switched by passing through the $C2/c$ phase, as it has a relatively smaller energy barrier. 

Also, we perform a comprehensive search for the low energy structures of $MX_2$O$_2$ ($M$=Mo and W; $X$=Cl and Br) compounds using the eigenvectors of $N_2^+$, $\Gamma_5^-$, and $M_5^-$ unstable phonon modes. The subgroup structures are generated by moving atoms according to the eigenvectors of single unstable mode and the combination of two and three unstable phonon modes. The results is tabulated in Table \textcolor{magenta}{S2} of the supporting information. The lowest-energy phase of all the $MX_2$O$_2$ compounds is $Cc$. Note that the recently reported $Fmm2$ phase of MoCl$_2$O$_2$ is formed by condensing the polar $\Gamma_5^-$ mode along (1,1) direction and is slightly higher in energy than the $Cc$ phase. The results is cross-checked by optB86b functional~\cite{PhysRevB.83.195131}. Our phonon calculation suggests $Fmm2$ phase is dynamically unstable and several lower energy phases are obtained by condensing the unstable phonon modes, see Figure~\textcolor{magenta}{S3} and Table~\textcolor{magenta}{S4} of the supporting information. Therefore, the experimentally observed $Fmm2$ of MoCl$_2$O$_2$ is a metastable phase. Note also that Lin {\it et al.} predicted that MoBr$_2$O$_2$ and WCl$_2$O$_2$ monolayers exhibit two-dimensional (2D) noncollinear ferrielectricity as a result of frustrated dipole order~\cite{PhysRevLett.123.067601}. Our results on MoBr$_2$O$_4$ octahedral distortion are consistent with their findings.

\hspace{0.4cm}

\noindent \textbf{Electronic structures and optical properties of MoBr$_2$O$_2$} \\
\noindent The electronic structures of the ground-state $Cc$ and reference $C2/c$ phase of MoBr$_2$O$_2$ exhibit significant differences, see Figure~\ref{bandstructure}. Although both phases have their valence band maximum (VBM) and conduction band minimum (CBM) located at the $\Gamma$ point of the first Brillouin zone, there is a notable distinction at the bottom of the conduction band. The presence of a flat conduction band (heavy electron) along the Y$_2$-$\Gamma$-M$_2$ direction, primarily originating from Mo 4$d_{yz}$ orbitals and exhibiting a significant mixture with Br 4$p$ orbitals, is lower in energy than that of the more dispersive band (light electron), which is derived mainly from $d_{xz}$ orbitals, in the $C2/c$ phase. Conversely, these two bands have an opposite ordering in the $Cc$ phase, leading to potential variations in electron conductivity and optical absorption. Furthermore, it is important to note that the $Cc$ phase exhibits distinct Rashba spin splitting at the VBM~\cite{rashba1960properties,PhysRevB.38.1806}, which arises from the broken inversion symmetry in its polar structure. In contrast, the centrosymmetric $C2/c$ phase does not display any Rashba spin splitting at the VBM. The evolution of Rashba spin splitting along the phase transition path from $Cc$ to $C2/c$ is depicted in Figure~\textcolor{magenta}{S4} of the supporting information. As the amplitude of the $\Gamma_2^-$ polar distortion decreases, the Rashba spin splitting gradually diminishes and eventually vanishes in the $C2/c$ phase, where the amplitude of the $\Gamma_2^-$ distortion is zero, which is similar to GeTe~\cite{https://doi.org/10.1002/adma.201203199}, Ag$_2$BiO$_3$~\cite{he2018tunable}, and organic–inorganic perovskite~\cite{wang2020switchable}. Due to the higher energy of the Br 4$p$ orbitals compared to the O 2$p$ orbitals, and the $d^0$ configuration of Mo$^{6+}$ cation, the valence band in MoBr$_2$O$_2$ is primarily composed of the Br 4$p$ orbitals, while the conduction band is mainly derived from the Mo 4$d$ orbitals. Interestingly, there is a noticeable hybridization between the Mo 4$d$ and Br 4$p$ orbitals at the bottom of the conduction band. The variation in the bandgap from the  $C2/c$ to $Cc$ phase is attributed to the alteration in the crystal field, which arises from the modifications in the Mo-O and Mo-Br bond lengths. For detailed information, refer to Figure~\textcolor{magenta}{S5} in the supporting material.

Remarkably, the bandgap of MoBr$_2$O$_2$ decreases by approximately 40 \% upon the transformation from the $Cc$ to the $C2/c$ phase, see the calculated bandgap using PBE, HSE06~\cite{10.1063/1.1564060}, and G$_0$W$_0$~\cite{PhysRev.139.A796} in the Table~\textcolor{magenta}{S3} of the supporting information. This significant disparity in bandgap between the polar phase and the reference phase is unusual in ferroelectric materials. For example, the bandgaps of the reference phase ($Pm\bar{3}m$) of BaTiO$_3$ and PbTiO$_3$ are nearly identical to those of the polar phase ($P4mm$), with only a few percentage points of discrepancy, see Table~\textcolor{magenta}{S3} of the supporting information. Our calculations on the changes in bandgap for PbTiO$_3$ align with the experimentally observed alteration in bandgap during the phase transition~\cite{vzelezny2015variation}. This can be primarily attributed to the deformation of the MoBr$_2$O$_4$ octahedral, leading to a modification in the crystal field splitting of Mo 4$d$ orbitals.

When an octahedral crystal field is applied, the $d$ orbitals of transition metals experience a division into two distinct groups, $t_{2g}$ and $e_g$. Nevertheless, due to the dissimilarity between the equator O$^{2-}$ and pole Br$^{-}$ in the homoleptic octahedron, the degeneracy of orbitals within $t_{2g}$ and $e_g$ can be further altered. In addition, the distortion of MoBr$_2$O$_4$ octahedron caused by the pseudo Jahn-Teller effect leads to further orbitals split. In the $C2/c$ phase, the $t_{2g}$ orbitals are arranged in increasing energy order as $d_{xz}$, $d_{yz}$, and $d_{xy}$. However, in the $Cc$ phase, the energy order of these orbitals changes to $d_{yz}$, $d_{xz}$, and $d_{xy}$. Therefore, the significant change of the bandgap between $Cc$ and $C2/c$ phases can be attributed to the crystal field splitting induced by polar distortion. MoCl$_2$O$_2$ exhibits an even more remarkable change in bandgap during the transition from the $Fmm2$ phase to the $I4/mmm$ phase, see Figure~\textcolor{magenta}{S6} of the supporting information.



\begin{figure}
	\centering
	\includegraphics[width=1.0\linewidth]{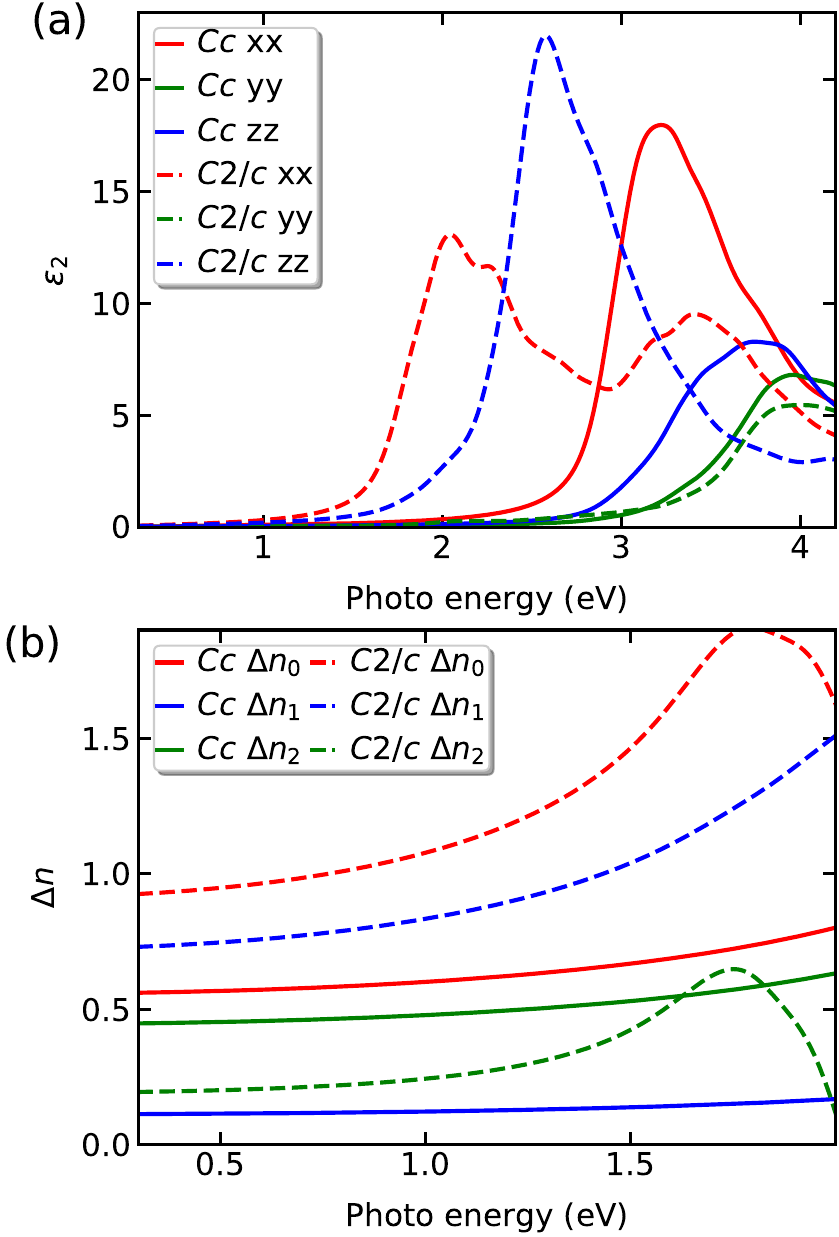}
	\caption{Optical properties of MoBr$_2$O$_2$ calculated using G$_0$W$_0$+IPA. (a) imaginary part of dielectric constant ($\epsilon_2$) of $Cc$ and $C2/c$ phases. (b) birefringence of $Cc$ and $C2/c$ phases.}
	\label{optics}
\end{figure}

The optical properties of $Cc$ and $C2/c$ phases of MoBr$_2$O$_2$ are calculated using the independent-particle approximation based on the many-body G$_0$W$_0$ eigenvalues. As shown in Figure~\ref{optics}, the optical absorption (imaginary part of dielectric constant $\epsilon_2$) of $C2/c$ and $Cc$ exhibit significant anisotropy, due to their low crystal symmetry. In the $Cc$ phase, the most intense absorption peak occurs at 3.2 eV in the $xx$ direction (the stacking direction), followed by peaks in the $zz$ and $yy$ directions (the corner-sharing octahedral directions) at higher photoenergies exceeding 3.8 eV. Conversely, in the $C2/c$ phase, the strongest absorption peak is observed in the $zz$ direction at 2.5 eV, followed by peaks in the $xx$ and $yy$ directions at 1.8 and 3.9 eV, respectively. As expected from the difference of the bandgaps, the shift of absorption peaks between $Cc$ and $C2/c$ is significant ($\sim$ 1 eV) for $xx$ and $zz$ directions. However, the absorption peak along $yy$ direction remains largely unaffected, suggesting a significant alteration of anisotropy. Additionally, the disparity in optical absorption between the $C2/c$ and $Cc$ phases can be attributed to the substantial difference in the sum of the squares of the dipole transition matrix elements (DTM$^2$)~\cite{doi:10.1021/acs.jpclett.7b01042} between the two phases. As shown in Figure~\ref{bandstructure}(c) and (d), the DTM$^2$ of these two phases are quite different. In the $Cc$ phase, the pairs of the highest valence band (HVB) to the lowest conduction band (LCB) and HVB-1 to LCB+2 have the largest DTM$^2$ values at the $\Gamma$ point than any other points. However, in the $C2/c$ phase, all the pairs have smaller DTM$^2$ at the $\Gamma$ point than at other points. The lines between the $\Gamma$ and Y$_2$ as well as that between the $\Gamma$ and V$_2$ have the largest DTM$^2$ values. Moreover, the birefringence of MoBr$_2$O$_2$ is strongly structure dependent due to the large disparity of anisotropy between these two phases: the $C2/c$ phase has much larger birefringence than $Cc$ at infrared red region, see Figure~\ref{optics}(b). Even the $C2/c$ phase shows remarkably larger birefringence than most compounds~\cite{feng2023visible}, which improves phase match of nonlinear optics. Lastly, the birefringence of MoBr$_2$O$_2$ is strongly correlated to its polar distortion, resulting in a significant difference between the $C2/c$ and $Cc$ phases. In the infrared region, the $C2/c$ phase exhibits significantly higher birefringence than the 
$Cc$ phase, as depicted in Figure~\ref{optics}(b). Remarkably, the $C2/c$ phase displays even greater birefringence than the majority of other compounds in literature~\cite{feng2023visible}, thereby enhancing the phase match for nonlinear optics.


\hspace{0.3cm}

\noindent \textbf{Tunable optical properties of MoBr$_2$O$_2$} \\
\noindent The substantial change in electronic structures and optical properties between $Cc$ and $C2/c$ phases allows us to manipulate and control these properties by modulating this phase transition. Given that the ground state of MoBr$_2$O$_2$ is a polar semiconductor, while the reference structure is a centrosymmetric semiconductor, it is plausible to manipulate the $Cc$ $\rightarrow$ $C2/c$ phase transition through three methods: (i) application of a strong enough external electric field, (ii) increasing the temperature, and (iii) above-bandgap optical excitation.

Due to the coupling between the dipole moment of the compound and the electric field, the polarization of a ferroelectric compound can be effectively switched by applying a sufficiently strong electric field, which causes the polar structure to transition from one direction to another, typically passing through the reference phase within a specific domain. Therefore, it is feasible to access the reference $C2/c$ phase if MoBr$_2$O$_2$ is a ferroelectric compound. While it is challenging to conclusively demonstrate the ferroelectricity of MoBr$_2$O$_2$ through DFT simulations alone, considering factors like domains that can affect switchability, we can utilize the energy difference ($\Delta{E}$) between the polar and reference phases, along with the polarization magnitude, as a rough estimation to determine if an applied electric field can overcome the energy barrier. Based on the calculated energy difference ($\Delta{E}$) and dipole moment of MoBr$_2$O$_2$, the estimated electric field is approximately 18 MV/m. This value is comparable to that of BaTiO$_3$, as shown in Figure~\textcolor{magenta}{S7} of the supporting information, suggesting that the polarization of MoBr$_2$O$_2$ can be switched by an external electric field. Hence, it is feasible to manipulate the electronic structures and optical characteristics of MoBr$_2$O$_2$ by applying an external electric field within a single ferroelectric domain. Note the single domain has been obtained in several ferroelectric materials, such as BaTiO$_3$~\cite{lee2021plane}. Since the $C2/c$ phase is metastable, it will go back to the $Cc$ phase when the external electric field is removed. Also, it is possible to achieve ultrafast tuning of visible light transmittance and infrared red birefringence in the single domain MoBr$_2$O$_2$ by applying a pulse electric field. This holds great promise for applications in information processing and energy harvesting.

To access the $C2/c$ phase, another approach is to increase the temperature of the material. This is possible due to the relatively small energy difference ($\Delta{E}$= 11 meV/f.u.) and polar mode amplitude ($Q$= 0.38 \AA) between the ground state $Cc$ and reference $C2/c$ phases, which are even smaller than those of PbTiO$_3$, as demonstrated in Figure~\textcolor{magenta}{S7} of the supplementary information. To prove the possible phase transition temperature, we conduct self-consistent phonon calculations for the $C2/c$ phase at finite temperatures, considering the contributions from both 3rd and 4th order force constants~\cite{doi:10.7566/JPSJ.87.041015,PhysRevLett.125.085901,PhysRevX.10.041029}. It turns out that the previously observed imaginary frequency $\Gamma_2^-$ completely disappears even at 300 K, see Figure~\textcolor{magenta}{S8} of the supplementary information. While the phase transition temperature is underestimated, this observation highlights the tendency for the $C2/c$ phase to be easily stabilized at the elevated temperatures. In addition, this finding suggests that temperature-induced changes have the potential to tune and manipulate the optical properties of MoBr$_2$O$_2$. Note that the phase transition induced by laser irradiation heat has been employed in 2D materials~\cite{doi:10.1126/science.aab3175}.

\begin{figure}
	\centering
	\includegraphics[width=1.0\linewidth]{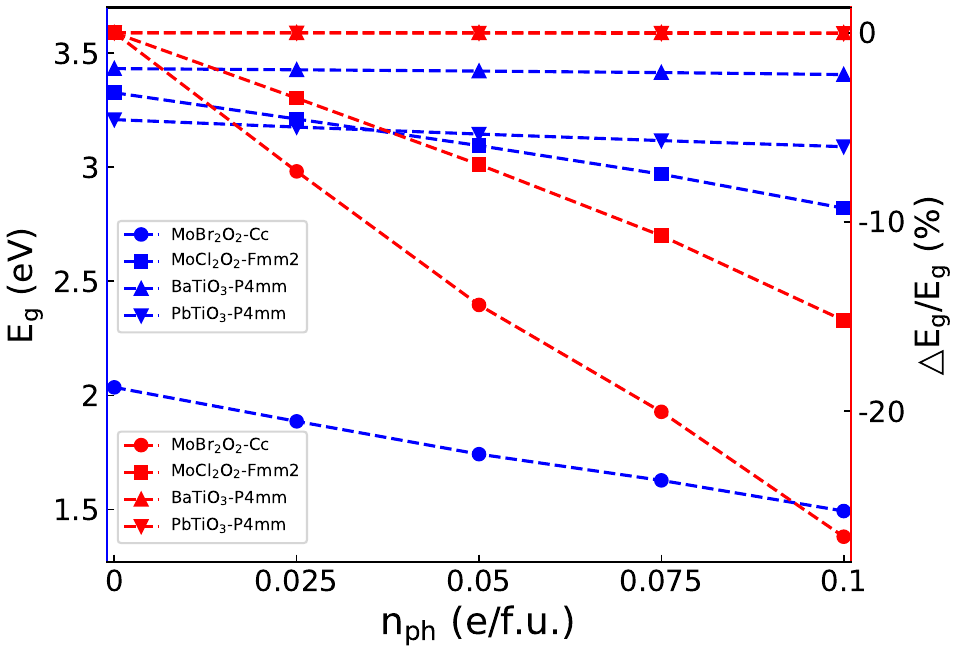}
	\caption{Bandgap changes as a function of photogenerated carrier ($n_{\rm ph}$) concentration.}
	\label{bandgapchangeligh}
\end{figure}

As mentioned above, when a ferroelectric semiconductor is subjected to light illumination with photon energies ($\hbar\omega$) above its bandgap ($E_g$), the generated carriers have the ability to screen its polarization and destabilize the polar phase. This scenario is reminiscent of the destabilization of the ferroelectric phase in BaTiO$_3$ due to carrier doping~\cite{PhysRevLett.104.147602,PhysRevLett.109.247601}. In this study, we carry out the simulation of photo-excited carriers by exciting electrons from valence bands to conduction bands. We optimize both the electron occupation and crystal structure concurrently, ensuring a comprehensive analysis of the system. Further details regarding this methodology can be found in the Method section. The suppression of photogenerated carriers towards polar distortion is illustrated in Figure~\textcolor{magenta}{S9}. Similar to BaTiO$_3$ and PbTiO$_3$, the magnitude of the $\Gamma_2^-$ distortion in the $C2/c$ MoBr$_2$O$_2$ structure decreases significantly as the number of photogenerated carriers ($n_{\rm ph}$) increases. The variation of the HSE06 bandgap with respect to $n_{\rm ph}$ is shown in Figure~\ref{bandgapchangeligh}. As the number of photogenerated carriers accumulates, the bandgap of $Cc$ MoBr$_2$O$_2$ exhibits a nearly linear decrease.

At the photogenerated carrier concentration of $n_{\rm ph}$ = 0.1 e/f.u., the relative reduction in bandgap ($\frac{\Delta{E_{\rm g}}}{E_{\rm g}}$) in $Cc$ MoBr$_2$O$_2$ is approaching 27 \%, which is significantly larger than that in the archetypal ferroelectric materials BaTiO$_3$ and PbTiO$_3$ ($<$ 1 \%). This can be attributed to the strong coupling between the bandgap and polar structure distortion in MoBr$_2$O$_2$. As depicted in Figure~\textcolor{magenta}{S6} of the supporting information, the larger $\frac{\Delta{E_{\rm g}}}{E_{\rm g}}$ of MoCl$_2$O$_2$ when the polar distortion is suppressed ($Q_{\Gamma_2^-}$=0) does not transfer to a larger $\frac{\Delta{E_{\rm g}}}{E_{\rm g}}$ under light illumination compared to MoBr$_2$O$_2$. This is due to the fact that the change in polar amplitude of MoCl$_2$O$_2$ with respect to $n_{\rm ph}$ is smaller than that of MoBr$_2$O$_2$. Note the carrier concentration of $n_{\rm ph}$=0.1 e/f.u. is approximately 10$^{21}$ cm$^{-3}$, which is achievable experimentally~\cite{PhysRevB.107.104109}. The nearly linear relationship between $\frac{\Delta{E_{\rm g}}}{E_{\rm g}}$ and $n_{\rm ph}$ indicates that a higher $n_{\rm ph}$ would result in a larger $\frac{\Delta{E_{\rm g}}}{E_{\rm g}}$.

\begin{figure}
	\centering
	\includegraphics[width=1.0\linewidth]{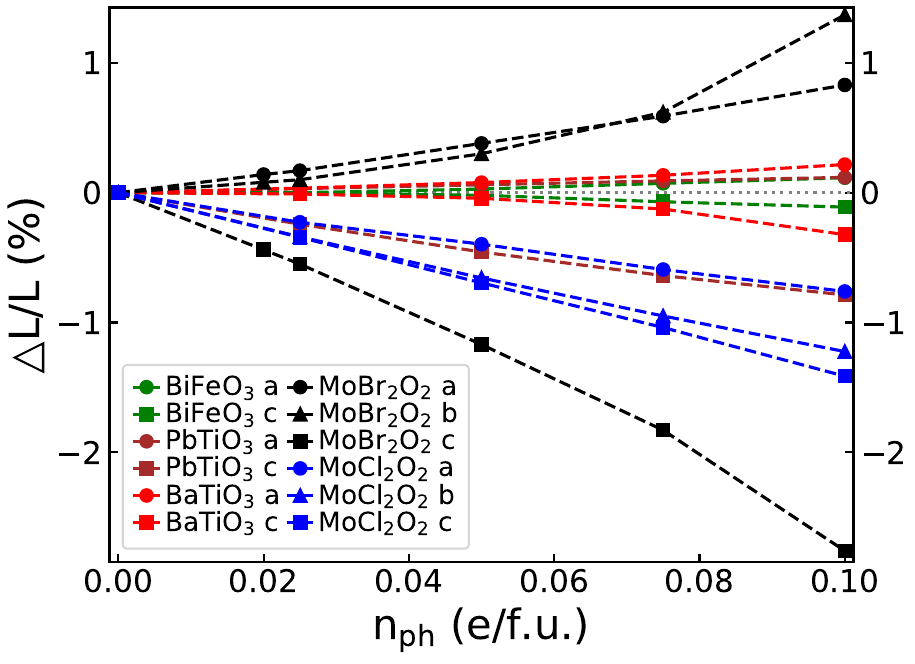}
	\caption{Relative change of lattice constants ($\Delta{\rm L}/{\rm L}$) as a function of photogenerated carrier ($n_{\rm ph}$) concentration. $R3c$ BiFeO$_3$ and $P4mm$ BaTiO$_3$ are used for comparison.}
	\label{Photostrictive}
\end{figure}

In addition to the tunable bandgap and optical properties observed under above-bandgap light illumination, MoBr$_2$O$_2$ also exhibits a giant photostrictive effect. This effect is characterized by a relative change in lattice constant ($\frac{\Delta{\rm L}}{\rm L}$) and is attributed to the converse piezoresponse and the screening of carriers generated by the photovoltaic effect on the dipole moment in ferroelectric compounds~\cite{10.1063/1.4905505,PhysRevLett.118.227401}. As shown in Figure~\ref{Photostrictive}, the lattice constants $a$ and $b$ of MoBr$_2$O$_2$ show a monotonic expansion, whereas the length of $c$ decreases more prominently as the $n_{\rm ph}$ increases. This behavior is understandable since the polar distortion (polarization) primarily occurs along the $c$ direction, while the reduction in volume under light illumination is alleviated by the expansion of $a$ and $b$. Remarkably, the magnitude of the lattice strain observed in MoBr$_2$O$_2$ is greater than that observed in BaTiO$_3$, PbTiO$_3$, and even BiFeO$_3$, which is a known compound that has large photostrictive effect\cite{kundys2010light,PhysRevLett.116.247401}. It is noteworthy that all three lattice constants of MoCl$_2$O$_2$  undergo a contraction upon light illumination, which starkly contrasts with the behavior observed in other compounds examined in this study and reported in previous works~\cite{PhysRevLett.118.227401,PhysRevLett.116.247401}. This unusual phenomenon is presumably because of a negative thermal expansion with elevating temperature in MoCl$_2$O$_2$. Materials exhibiting a substantial photostrictive effect hold great promise for applications in actuating and photoacoustic devices, light-driven relays, light-driven motors, photostrictive sensors, and numerous other applications~\cite{https://doi.org/10.1002/adfm.202010706}.

In conclusion, our study presents compelling evidence for the presence of electric-field switchable polarization and an ultrastrong coupling between polar distortion and electronic structures as well as optical properties in the layered compound MoBr$_2$O$_2$, utilizing first principles calculations. These distinct characteristics position MoBr$_2$O$_2$ as an exceptional platform for achieving tunable Rashba spin-splitting, electrochromism, thermochromism, photochromism, and nonlinear optics by applying an external electric field, increasing temperature, and intense light illumination. Furthermore, an exceptionally large photostrictive effect has been observed in MoBr$_2$O$_2$. These discoveries not only demonstrate the feasibility of achieving cross-control through the coupling of multiple order parameters within a single material but also establish the foundation for comprehensive control over the properties of materials with various applications in actuator technology, smart windows, sensor technologies, and information technology, enabling ultrafast property tuning.

\section{Methods}
The Majority of the DFT calculations in this work are performed using the Vienna {\sl ab initio} Simulation Package (VASP)~\cite{vasp1,vasp2}. The projector augmented wave (PAW~\cite{PAW1,PAW2}) pseudo potential, plane wave basis set, and SCAN+rvv10~\cite{PhysRevX.6.041005} exchange-correlation functional were used. The $\Gamma$-centered $k$-point grids with a density of more than 8000 $k$-points per reciprocal atom (KPPRA) were used to sample the Brillouin zone. All the structures are fully relaxed until the force on each atom is smaller than 0.01 eV/\AA.
Second-order force constants were computed by using the finite displacement method as implemented in the \texttt{phonopy} package\cite{phonopy}. The bandgap is calculated using HSE06 functional\cite{10.1063/1.1564060} and optical properties were computed using independent particle approximation (IPA) based on the energy eigenvalues calculated using many-body G$_0$W$_0$~\cite{PhysRev.139.A796, PhysRevB.74.035101,PhysRevB.75.235102}.
The G$_0$W$_0$ calculations employed the GW versions of the PAW potentials, with semicore electrons included where available. A frequency grid containing 380 points, an energy cutoff on the plane-wave basis of 500 eV (together with a cutoff for the response function of 300 eV) were used; 1280 unoccupied orbitals were included in the self-energy band summation. The Quasi-Particle corrections were determined on a $6 \times 6 \times 6$ \textit{k}-point mesh and interpolated to the  $16 \times 16 \times 16$ mesh employed for computing the optical properties. Symmetry analysis was performed by using ISOTROPY\cite{isotropy} and Bilbao~\cite{Orobengoa:ks5225,Perez-Mato:sh5107}.

The structure and properties calculations under light illumination is computed using extended version of ABINIT code\cite{GONZE20092582} by Paillard~\cite{PhysRevLett.116.247401}. The photo-excited thermalized carriers are mimicked using Fermi-Dirac distribution with two quasi-Fermi-levels $\mu_{e}$ and $\mu_{h}$. During our DFT calculations, the density is self-consistently converged under the constraint of having $n_e$ = $n_{ph}$ (and $n_h$ = $n_{ph}$) electrons (holes) in the CB (VB). All structures considered in this work were geometrically relaxed under the photo-excitation constraint mentioned above.

\hspace{0.5cm}

\bibliography{ref}
\bibliographystyle{aipnum4-1}
%
%
\section{Acknowledgments}
Z.L, Y.Y, and J.H. acknowledge the support of the Fundamental Research Funds for the Central Universities China (USTB). L.V. and C.F. acknowledge computing time at the Vienna Scientific Cluster. The computing resource was supported by USTB MatCom of Beijing Advanced Innovation Center for Materials Genome Engineering.

\hspace{0.2cm}

\textbf{Corresponding author}
Correspondence to ylyang@ustb.edu.cn, \\
jghe2021@ustb.edu.cn

\end{document}